August 21, 2005

**On the Electrodynamics of Cosmic Repulsion**

Howard D. Greyber
2475 St. Lawrence Dr. San Jose, CA 95124, U.S.A.          hgreyber@yahoo.com

Subject:    Physics - - - Astrophysics

**Abstract:**
Two groups of astronomical observers, one headed by Saul Perlmutter and the other headed by Robert Kirshner, recently found an amazing transition, from the expected slowing down of the expansion of the Universe due to Gravity, to an accelerating expansion of the Universe beginning at an age of the Universe of about nine billion years. Profound questions that arise are: What is this "dark energy" causing this result? Why has it started to overcome attractive gravity only in the last five billion years of our 14 billion year old Universe? My answers are based on the "Strong" Magnetic Field model (SMF) which uses both Gravity and the Physics of Plasmas. SMF has previously been used to explain the nature of the AGN/Quasar central engine, the evolution of galaxies, quasars and jets, the origin of large-scale magnetic fields, and the large-scale structure of galaxies in our Big Bang Universe (1). The astronomical observations of the present-day acceleration of the expansion of the Universe are explained by adding well-known facts about Albert Einstein's general relativity equations, together with the SMF model of our Early Universe, to explain the observed Cosmic Repulsion.

Body of Article:
To help understand the physics of this surprising result of a present-day accelerating expansion of the Universe, I excerpt a few sentences from the Astrophysics/Cosmology section of a website by cosmologist Stephen Hawking (2):
   "As you can see from the Einstein equations, such accelerating expansion, or inflation as it was called, required either negative energy, or negative pressure. One gets into a lot of trouble if one allows negative energy. One would get runaway creation of particle pairs, one with positive energy and the other with negative. But there is no reason to rule out negative pressure. That is just tension, which is a very common condition in the modern world. - - The original model of inflation , which came to be known as old inflation, had various problems. How did the universe get into a false vacuum state in the first place, and how did it get out again? - - Andrei's (Linde) idea removed the need that the Universe began in a false vacuum. However one needed to explain why the field should have been nearly constant over a region, with a value that was not at the minimum of the potential."

This research is based on my SMF model, described in the article, "On the Electrodynamics of the Big Bang Universe", which is in turn expanded from a rather considerable body of earlier research by me since 1961 (1). Cosmic repulsion is only mentioned in a brief Appendix in my recent "Texas At Stanford" article. Up to then, my basic purpose was to explain the origin of large-scale astrophysical magnetic fields, the origin of the large-scale structure of galaxies and quasars, and the origin of the central engine, morphology, dynamics, jets and energetics of galaxies and quasars. Applying the well-known physics of plasmas, a "Strong" Magnetic Field model (SMF) was created that explains the origin of a very large-scale magnetic field at the boundaries of each supercluster, and simultaneously explain the observed large-scale structure of galaxies. Present-day models using only Gravity plus ad hoc assumptions, thus ignoring the known facts of plasma physics, are inadequate.

SMF involves both Gravitation and Cosmic Magnetism (i.e. using the Electromotive Force which is immensely stronger than Gravity and is also long-range). The unique SMF processes act mainly, but not entirely, during the era of Combination Time (sometimes called Decoupling Time, when the radiation is able to escape from the matter). That means from an age of 300,000 years to, perhaps, a million years in the Big Bang Universe. In SMF, the word "primordial" means processes acting at and after Combination Time, as George Gamow used it. The word "strong" has been used by me since 1961 in precisely the same sense that the outstanding physicist, Jacob Zel'dovich, used it in 1983, "A major challenge is to understand strong magnetic fields whose energies greatly exceed those of hydrodynamic motions" (3).

Valerie DeLapparent, who had worked on the first large-scale survey of galaxies with Margeret Geller and John Huchra at Harvard, describes the observed large-scale galaxy distribution as "sharp sheet-like structure delineating voids"(4). Stephen Gregory concluded "the large-scale structure we observe today had to be impressed somehow onto the proto-galactic material; before it collected into galaxies" (5). In carefully reasoned conclusions, Jaan Einasto, Alexei Starobinsky and eight other scientists wrote "We present evidence for a quasi-regular three dimensional network of superclusters and voids, with the regions of high density separated by 120 Mpc. If this describes the distribution of all matter (luminous and dark), then there must exist some hitherto unknown process that produces regular structure on large-scales".(6) In my opinion, SMF is their "hitherto unknown process".

Exploiting facts about Spinodal Decomposition, SMF describes how just after Combination Time, the Universe exists in patches of various size volumes. Then SMF explains how, at Combination Time, a growing electric current is created along the fully ionized plasma (FIP) boundaries, due to the huge flows of photons from the surrounding very slightly ionized plasma patches (SIP) onto the borders of each patch of FIP. The increasing current causes the well-known Pinch Effect to pull in charged particles from the surrounding plasma, beginning to create the void. Then, the gravitational attraction of the gradually increasing mass density along the sharp boundaries of the Fully Ionized Patch (FIP) acts to pull in *both* neutral and charged particles into these sharp boundaries, thus creating a high vacuum at the center of each void. The increasing mass density causes critical density to be reached in many local volumes along the boundaries, a couple hundred million years later, of what we observe and call today a Supercluster of galaxies.

Thus, gravitational collapse of giant plasma clouds *in the presence of the primordial magnetic field* will form quasars, galaxies and stars around each Supercluster void. However, because the Pinch Effect is known to be unstable, gravitational collapse will occur only at certain places along the Supercluster boundaries, while leaving dark matter and ordinary plasma matter present all along these boundaries.

My SMF Model provides astrophysics with a *wholly new classification of galaxies*, in addition to that created by Edwin Hubble, i.e. the morphology, dynamics, jets and energetics of objects of galactic dimension are classified and determined by the Ratio of magnetic energy to rotational energy in the particular object. The ratio is highest for quasars, BL Lacs and blazars, decreasing steadily for giant elliptical radio galaxies, then less for Seyfert and Markarian galaxies, is low for spiral galaxies and even lower for ordinary featureless elliptical galaxies. However the activity observed is also a function of the matter accretion rate at the time observed.

In a new article, astro-ph/0507274, "Mapping Extreme-scale Alignments of Quasar Polarization Vectors", Damien Hutsemekers et al, again confirm that optical quasar polarization vectors are *not* randomly oriented over the sky, with a probability often in excess of 99.9%. Probably, in accord with SMF, the quasar magnetic structural axes are themselves coherently oriented (and thus aligned) by being formed closer (than other types of galaxies) to the SMF currents in the boundaries. In SMF, quasars are formed in the comparatively strong magnetic field and relatively strong density close to the boundaries of the huge DeVaucouleurs Superclusters that are generated by the SMF processes. It would be helpful if a search were made for a similar alignment of radio quasar polarization vectors over huge separation distances.

For background on Albert Einstein's remarkable equations for general relativity and on the expanding cosmos, one can peruse books by Alan Guth, Brian Greene, Andrei Linde, Stephen Hawking, Scott Dodelson and others. As Brian Greene remarks, it is not difficult to see how an accelerated expansion can arise (7). Greene and Dodelson show that one of Einstein's equations is

$$(1/a)\, d^2a/dt^2 \;=\; (-4\pi/3)G\,(3(\rho + 3p)) \qquad (1)$$

where $a$ is the scale factor, $\rho$ is the energy density, and $p$ is the pressure density. One observes that if the right hand side of the equation is positive, the scale factor for that region will grow at an accelerating rate, meaning the rate of growth of the Universe will accelerate with time. However, a simpler direct result of Einstein's equations is

$$p \;<\; -\rho/3 \qquad (2)$$

Einstein's general relativity cosmological term, lambda, implies that while positive pressure adds to attractive gravity, negative pressure contributes to "negative gravity", or a repulsive gravity. That upsets Newtonian physics where gravity is always attractive. In general relativity, gravity and pressure are separate concepts. Pressure, like mass and energy, is a source of gravity. But if a region has a negative pressure, it contributes a *gravitational push* to the gravitational field of that region, not a pull. Einstein found that the strength of the repulsive gravitational force is cumulative since *a larger region of negative pressure means more outward pushing.* Einstein also discovered that the repulsive gravitational force is important *only over huge cosmological expanses,* certainly not on the scale of our Solar System.

The lambda term appears to act like a vacuum energy density, which is popular in particle physics research because a type of vacuum energy density is used in their Higgs mechanism for spontaneous symmetry breaking. In particle physics, the grand unified theories (GUT) are described by using the ubiquitous Higgs fields. The energy density of the vacuum remains constant as the vacuum grows in size with an expanding universe. Since the energy density is always positive but close to zero in a vacuum, the vacuum pressure, p, is always negative, representing a negative pressure and of course a repulsive gravity force.

The famous general relativity theoretical physicist, Joseph Weber, who is better known for being the first physicist to carry out experiments to detect gravitational waves, produced some results relevant here (8) Using a clever mathematical transformation of Einstein's equations, Weber deduced, assuming that the gravitational force extends beyond the radius of galactic clusters, that

$$\text{lambda is} \ < \ 10^{-52} \ \text{cm}^2 \qquad \text{and the graviton mass is} \ < \ 10^{-63} \ \text{grams}$$

Clearly, the SMF processes leave a huge volume patch of extremely high vacuum, extraordinarily far out of equilibrium, that was first termed FIP (fully ionized plasma), at the center of each void inside each Supercluster. Notice that this is occurs about 46 orders of magnitude in time *later* than the time of Inflation. Alan Guth uses the term "false" vacuum, which has the same gravitational repulsion as the term Albert Einstein introduced into his equations for general relativity, and was removed a decade later. Guth says that "false" in this context simply means "temporary". The Einstein term has various names, lambda, cosmological constant, negative pressure, dark energy, vacuum energy, quintessence, etc. The commonly used term "supercooling" is deceptive, really describing in general relativity one mechanism for obtaining a volume of very high vacuum very far from equilibrium.

For SMF, it is intriguing that recently a group of theorists developed a new way to represent the CMB fluctuations, observed by WMAP, in terms of vectors. They revealed some unexpected correlations in 2004, and that one vector lies very near the plane defined by the local Supercluster of galaxies, termed the supergalactic plane. In a very recent article by Starkman and Schwarz (9), they report in some detail, stating "mysterious discrepancies have arisen between theory and observations of the "music" of the cosmic microwave background". They also say "The results could send us back to the drawing board about the early universe", which is just what the "Strong" Magnetic Field (SMF) model does.

**Conclusions:**
SMF demonstrates that each huge FIP vacuum patch is extremely far out of equilibrium, and represents a negative pressure, and thus a source of cosmic repulsion. If our Big Bang Universe, infinitely far beyond our own Supercluster, has similar Superclusters with huge patches of very high vacuum inside each of the voids everywhere in the Universe, as DeLapparent and Einasto, Starobinsky et al deduce from the astronomical observations, then this explains the observed accelerating expansion of the Universe as being due to Cosmic Repulsion caused by the unique SMF processes acting in every Supercluster in the Universe.

The reason that Cosmic Repulsion overcame the effect of Gravity only at a Big Bang age of about nine billion years is that right after Combination Time, each Supercluster, its void and its high vacuum region are quite small. Thus, at that very early time, the Cosmic Repulsion was very small. But, the continuing expansion of the Universe increases the volume of each void over billions of years and dramatically increases the volume of the region with extremely high vacuum that is inside each void. Thus the Cosmic Repulsion eventually caught up and overcame the slowly decreasing effect of attractive Gravity.

It is important to note that over the next few billion years, that even as the effect of Gravity slows down as the expansion of the universe increases, the Cosmic Repulsion from the growing huge extremely high vacuum FIP at the center of each cosmic void will increase. However, if an unknown mechanism exists, somehow transferring a sufficiently high rate of matter into each very high vacuum region in each void in the Universe, conceivably the accelerating expansion might be slowed. What wins out in the future is unknown.

**Acknowledgements**:
   It is a pleasure to thank the late Donald Howard Menzel for friendly sage advice and warm encouragement, and the great John Archibald Wheeler for friendly insightful comments.